\documentclass[12pt, preprint]{iopart}
\usepackage{graphicx}
\usepackage{color}
\begin{document}

\title{An approach to quantum transport based on reduced hierarchy equations of motion: Application to a resonant tunneling diode}

\author{Atsunori Sakurai and Yoshitaka Tanimura}
\address{Department of Chemistry, Graduate School of Science, Kyoto University, Kyoto 606-8502, Japan}
\ead{sakurai@kuchem.kyoto-u.ac.jp and tanimura@kuchem.kyoto-u.ac.jp}
\begin{abstract}
The quantum dissipative dynamics of a tunneling process through double barrier structures is investigated on the basis of a rigorous treatment for the first time. We employ a Caldeira-Leggett Hamiltonian with an effective potential calculated self-consistently, accounting for the electron distribution. With this Hamiltonian, we use the reduced hierarchy equations of motion in the Wigner space representation to study the effects of non-Markovian thermal fluctuations and dissipation at finite temperature in a rigorous manner. Hysteresis, double plateau-like behavior, and self-excited current oscillation are observed in a negative differential resistance (NDR) region of the current-voltage curve. We find that while most of the current oscillations decay in time in the NDR region, there is a steady oscillation characterized by a tornado-like rotation in the Wigner space in the upper plateau of the NDR region.
\end{abstract}

\maketitle
The Caldeira-Leggett (or Brownian) Hamiltonian has been applied to the investigation of quantum dissipative dynamics in several fundamental contexts, including quantum tunneling\cite{CLAnnlPhys1983,CLPhysica1983,WaxmanLeggett1985}, chemical reactions\cite{Miller1989},  SQUID rings\cite{Chen1986}, nonlinear optical response\cite{MUKAMEL95}, and quantum ratchets\cite{Hanggi09}.
Because a complete model of quantum dissipative dynamics must treat phenomena that can only be described in real time, a great deal of effort has been dedicated to the problem of numerically integrating equations of motion derived from the Hamiltonian that describe real-time behavior\cite{Waxman85,Coffey07,Jyoti2011}. Although such equations are analogous to the classical kinetic equations, which have proved to be useful in
the study of classical transport phenomena, they are difficult to derive in a quantum mechanical framework without approximations and/or assumptions. 
In this paper, we demonstrate that the reduced hierarchy equations of motion (HEOM) in the Wigner space representation provide a powerful method to study quantum dissipative dynamics in systems subject to non-Markovian and non-perturbative thermal fluctuations and dissipation at finite temperature\cite{TKJSPS1989,Tanimura91, TJPSJ2006}.
As an example, we employ a model describing the thermal effects in resonant tunneling diodes (RTDs)\cite{Datta, FerryGoodnickBird}.

Due to quantum effects, an RTD system exhibits novel negative differential resistance (NDR) in its current-voltage (I-V) relation\cite{CEsakiTsuAPL1974}. Moreover, current oscillations\cite{SollenerAPL1984}, plateau-like behavior and hysteresis of the I-V curve have been observed in the NDR region\cite{GoldmanTsuiPRL1987}. 
Theoretically, Frensley found NDR in the I-V curve in a numerical computation treating a quantum Liouville equation in the Wigner representation that ignored phonon-scattering processes\cite{FrensleyPRB1987, FrensleyRevModPhys1990}. Kluksdahl et al. incorporated dissipative and self-consistent effects by employing the Poisson-Boltzmann equation
by adopting a relaxation time approximation and succeeded in modeling the experimentally 
observed hysteresis behavior of the I-V curve\cite{FerryPRB1989}.
Jensen and Buot developed a numerical scheme to treat systems of the same kind
and reported that current oscillation and plateau-like behavior arise from intrinsic bistability\cite{BuotPRL1991,BuotJAP2000}. 

In the previous studies, dissipative effects of electron transport have been taken into account by using the Boltzmann equation. The quantum version of the Boltzmann equation, however, employs several approximations and assumptions, and for this reason, the validity of that equation in treating quantum dissipative dynamics is questionable. We wish to investigate dissipative dynamics in a more quantum mechanically rigorous manner. With this aim, we employ a model based on the Caldeira-Leggett (or Brownian) Hamiltonian\cite{CLPhysica1983,CLAnnlPhys1983},
\begin{equation}
\hat H = \frac{{{{\hat p}^2}}}{{2m}} + U(\hat q;\;t) + \sum\limits_j {\left[ {\frac{{\hat p_j^2}}{{2{m_j}}} + \frac{{{m_j}\omega _j^2}}{2}{{\left( {{{\hat x}_j} - \frac{{{c_j}\hat q}}{{{m_j}\omega _j^2}}} \right)}^2}} \right].}
\label{eq:CLHamiltonian}
\end{equation}
Here, $m$, $\hat{p}$ and $\hat q$ are the mass, momentum and position
variables of an electron, and $m_j$, $\hat{p}_j$, $\hat x_j$ and $\omega_j$ are the mass, momentum, position and frequency variables of the $j$th phonon oscillator mode. The quantities $c_j$ are coefficients that depend on the nature of the electron-phonon coupling.
Although the validity of the above Hamiltonian for describing electron transport phenomena has not yet been investigated thoroughly, because it has a firm quantum mechanical foundation, and because we feel that it is necessary to investigate the validity of the previous theoretical results in a fully quantum mechanical treatment, we believe that an investigation employing eq.(1) will be helpful in constructing a general understanding of electron transport phenomena. 

In eq.(1), $U(\hat{q},t)$ is an effective potential for an electron. It can be written $U(\hat q;\;t) = {U_{static}}(\hat q) + {U_{self}}(\hat q;\;t)$, where ${U_{static}}(\hat q)$ is the static part, determined from the conduction band structure, and ${U_{self}}(\hat q;\;t)$ is the self-consistent part, calculated from the electron distribution using the Poisson equation. 
The heat bath can be characterized by the spectral distribution function, defined by $J(\omega ) = \sum\nolimits_j {c_j^2\delta (\omega  - {\omega _j})/(2{m_j}{\omega _j})}$, and the inverse temperature, $\beta  = 1/kT$, where $k$ is the Boltzmann constant. 
We assume the Drude distribution, given by $J(\omega ) = \left( {m\zeta /\pi } \right)\left( {{\gamma ^2}\omega /({\gamma ^2} + {\omega ^2})} \right)$, where the constant $\gamma $ represents the width of the spectral distribution of the collective phonon modes and is the reciprocal of the correlation time of the noise induced by phonons. The parameter $\zeta $ is related to the electron-phonon coupling strength. 
We then employ the reduced hierarchy equations of motion (HEOM) approach, which can be used to treat non-Markovian and non-perturbative system-bath interactions at finite temperature with no approximation\cite{TJPSJ2006}.
In the HEOM formalism, the reduced density operator is expressed in terms of the auxiliary hierarchy density matrix elements, $\rho _{{j_1}, \cdots ,{j_K}}^{(n)}$, where the indices $n$ and $j_k$ arise from the hierarchal expansion of the decay functions $e^{- \gamma t}$ and $e^{- \nu_k t}$, where $\nu_k = 2 \pi k /\beta \hbar$ is
the $k$th Matsubara frequency.\cite{TJPSJ2006,ITJSPS2005} Then the 0th element is identical to $\hat \rho _{0,0, \cdots ,0}^{(0)}(t) = {\rm{T}}{{\rm{r}}_B}\left\{ {{{\hat \rho }_{{\rm{tot}}}}(t)} \right\}$. 
The HEOM has been used to study chemical reactions\cite{Tanimura91, 
shi09}, linear and nonlinear spectroscopy\cite{STJPCA2011,Yan12}, exciton and electron transfer\cite{IshiFlem09,Kramer12,Schulten12,TTJPSJ08,TJCP2012}, and quantum information\cite{Dijkstra10,Nori12}.
The HEOM is ideal for studying quantum transport systems when we employ the Wigner representation,
because it allows us to treat continuous systems utilizing open boundary conditions and periodic boundary conditions\cite{Tanimura91,FrensleyRevModPhys1990}.

The equations of motion are written in hierarchical form as follows\cite{TJPSJ2006,STJPCA2011}:
\begin{equation}
\begin{array}{l}
\frac{\partial}{\partial t} W_{{j_1}, \cdots ,{j_K}}^{(n)}(t) \\ =  - \left[ {{{\hat L}_{qm}} + \hat \Xi ' + n\gamma  + \sum\limits_{k = 1}^K {{j_k}{\nu _k}} } \right]W_{{j_1}, \cdots ,{j_K}}^{(n)}(t)\\
 + \hat \Phi \left[ {W_{{j_1}, \cdots ,{j_K}}^{(n + 1)}(t) + \sum\limits_{k = 1}^K {W_{{j_1}, \cdots ,({j_k} + 1), \cdots {j_K}}^{(n)}(t)} } \right]\\
 + n\gamma {{\hat \Theta }_0}W_{{j_1}, \cdots ,{j_K}}^{(n - 1)}(t) + \sum\limits_{k = 1}^K {{j_k}{\nu _k}{{\hat \Theta }_k}W_{{j_1}, \cdots ,({j_k} - 1), \cdots {j_K}}^{(n)}(t).} 
\end{array}
\end{equation}
Here, the quantum Liouvillian in the Wigner representation is defined by
$-{\hat L_{qm}}W(p,q) =- \frac{p}{m}\frac{\partial }{{\partial q}}W(p,q)-\int_{ - \infty }^\infty  {\frac{dp'}{2\pi \hbar^2 }} U_{\rm{w}} (p-p',q;\;t)W(p',q)$ with ${U_{\rm{w}}}(p,q) = 2\int_0^\infty  {dr} \sin \left( {\frac{{pr}}{\hbar }} \right)\left[ {U\left( {q + \frac{r}{2}} \right) - U\left( {q - \frac{r}{2}} \right)} \right]$\cite{FrensleyRevModPhys1990}.
The other operators appearing in eq.(2) are defined as $\hat \Phi  = \frac{\partial}{\partial p}$, ${\hat \Theta _0} = \zeta \left[ {p + \frac{{{m} \hbar \gamma }}{2}\cot \left( {\frac{{\beta \hbar \gamma }}{2}} \right){\frac{\partial}{\partial p}}} \right]$, ${\hat \Theta _k} = \frac{2m\gamma^2\zeta }{\beta(\nu_k^2-\gamma^2)}\frac{\partial}{\partial p}$, and $\hat \Xi ' =  - \frac{{m\zeta }}{\beta }\left[ {1 - \frac{{\beta \hbar \gamma }}{2}\cot \left( {\frac{{\beta \hbar \gamma }}{2}} \right) - \sum\nolimits_{k = 1}^K {\frac{2 \gamma^2}{\nu_k^2-\gamma^2}} } \right]\frac{\partial ^2}{\partial {p^2}}$. If the combinations of $n$ and $j_k$ are sufficiently large that the condition $N \equiv n + \sum\nolimits_{k = 1}^K {{j_k} \gg {\omega _c}/\min (\gamma ,\nu_1 )}$ holds, where ${\omega _c}$ is the characteristic time scale of electron motion, the infinite hierarchy in eq.(2) can be truncated at those values of $n$ and $j_k$ with negligible error by setting $\dot W_{{j_1}, \cdots ,{j_K}}^{(n)}(t) =  - \left( {{{\hat L}_{qm}} + \hat \Xi '} \right)W_{{j_1}, \cdots ,{j_K}}^{(n)}(t)$\cite{ITJSPS2005,STJPCA2011}.  

\begin{figure}[t]
 \includegraphics[width=7cm]{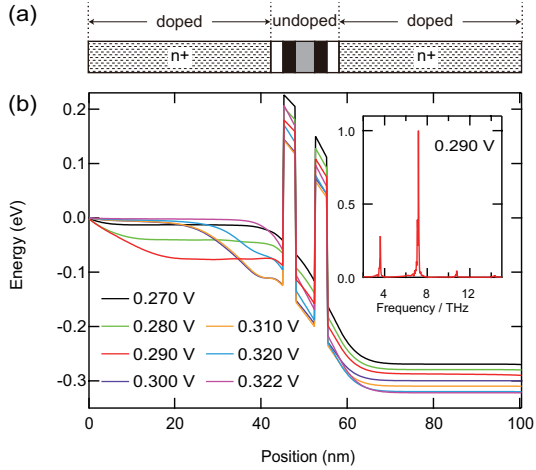}
\caption{\label{fig1:g} (a) The structure of the RTD. The contact regions are GaAs doped 2$\times 10^{18}\mathrm{cm}^{-3}$ (42.375 nm), the spacer layers are undoped GaAs (2.825 nm), the barriers are undoped AlGaAs (2.825 nm), and the well is undoped GaAs (4.52 nm). (b) The averaged potential $\bar U(q)$ for various values of the bias voltage. The potential on the emitter side is flat when the biases are 0.270V and 0.322V. 
Contrastingly, for bias voltages in the NDR region, a basin is formed on the emitter side. This basin becomes broad for 0.290V. The basins are deepest when the biases are 0.300V and 0.310V, corresponding to the lower plateau region of the I-V curve in Fig. 2 (a). The inset plots the Fourier components of the current oscillation for 0.290V.}
\end{figure}

We now discuss our numerical simulations using the model described above. As the static potential, we employed a double-barrier structure, which models a hetero-structure of GaAs sandwiched between two thin AlGaAs layers. The conduction band edge consists of a single quantum well bounded by tunneling barriers. The effective mass of an electron was assumed to be constant across the device and equal to $0.067{m_0}$, where $m_0$ is the electron mass in vacuum. 
The size of the position momentum space is 100nm$\times$0.704nm$^{-1}$, which was discretized onto a 356$\times$200 lattice. 
The structure of the RTD is illustrated in Fig. 1 (a). The potential barriers are 0.27eV. 
We set the parameters used in the HEOM as $\gamma  = 12.1$THz (${\gamma ^{ - 1}} = 8.27$fs), $\zeta  = 72.5$GHz (${\zeta ^{ - 1}} = 13.8$ps), and $T$=300K to create conditions close to those used in previous theoretical studies\cite{FrensleyPRB1987,FerryPRB1989,FrensleyRevModPhys1990,BuotPRL1991,BuotJAP2000}. The depth of the hierarchy and the number of Matsubara frequencies were chosen as $N \in \{2, 3, 4, 5, 6 \}$ and $K \in \{1, 2, 3 \}$, 
respectively. The third-order up-wind and down-wind difference schemes respectively were used in the positive and negative $p$ regions to approximate the spatial derivative of the kinetic term, $ - (p/m)\partial /\partial q$, in order to facilitate implementation of the inflow and outflow boundary conditions\cite{FrensleyPRB1987,FrensleyRevModPhys1990}. 
Fourth-order centered difference schemes were employed for the other derivatives of $p$. 
The fourth-order Runge-Kutta method was used for the time evolution.
The inflow boundary conditions are set by stipulating
$W^{(n)}_{j_1, \cdots , j_k} (p<0, q =L)$ and
$W^{(n)}_{j_1, \cdots , j_k} (p>0, q=0)$
to be given by the equilibrium distribution of a free particle
calculated from the HEOM with periodic boundary conditions\cite{Tanimura91}.
Due to fluctuations and dissipation, the flow of a wavepacket reaches a steady state even when there exists bias voltage. The validity of the boundary conditions was verified by considering several system sizes. 
At each step, a self-consistent potential ${U_{self}}(q;t) =  - e \phi (q;t)$ was calculated using the Poisson equation $- \varepsilon \frac{\partial ^2}{\partial {q^2}}\phi (q;t) = e\left[ {{n^ + }(q) - P(q;t)} \right]$, where $\varepsilon$ is the dielectric constant ($\varepsilon  = 12.85$), ${n^ + }(q)$ is the doping density (${n^ + } = 2 \times {10^{18}}{{\mathop{\rm cm}\nolimits} ^{ -3}}$ in the doped region), and $P(q;t) = \int_{ - \infty }^\infty  {dp} W_{0,...,0}^{(0)}(p,q;t)/2\pi \hbar $ is the electron density calculated from the Wigner distribution.

\begin{figure}[t]
 \includegraphics[width=8.5cm]{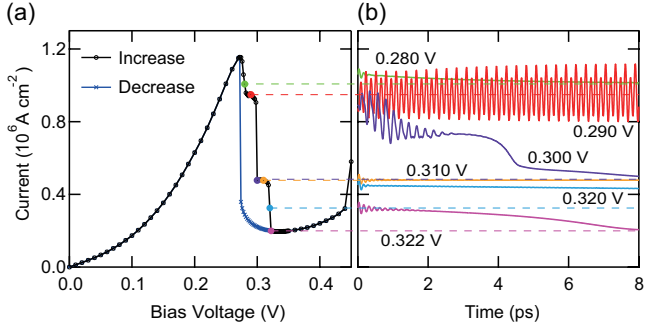}
\caption{\label{fig2:g} (a) Steady current as a function of applied bias voltage. 
The black curve represents the case in which the bias is increased, and the blue curve 
represents the case in which the bias is decreased. 
(b) Time evolution of the current for various bias voltages in the NDR region. 
A non-decaying current oscillation is observed in the upper plateau region between 0.284V and 0.298V.
}
\end{figure}

We calculated the current-voltage (I-V) characteristics according to the following procedure. Under the inflow boundary conditions specified above, we integrated eq.(2) with the effective potential $U(\hat q;\;t)$
evaluated iteratively using the Poisson equation at zero bias voltage. 
When the distribution reached the steady state, the current was calculated using the 
distribution $\lim_{\rm t \rightarrow \infty} W_{0, \cdots, 0}^{(0)} (p,q;t)$, and then the final distribution was used as the initial distribution in the next bias step. 
We increased the bias from 0.000V to 0.450V and then decreased it to 0.000V with bias steps of 0.01V and 0.002V in the normal and NDR regions, respectively.
At each step, we integrated the equations of motion until the system reached the steady state distribution. The time periods of the integrations were typically 2-4ps in the normal region and 15-40ps in the NDR region.
The calculated effective potentials (${\bar U}(q) = \lim_{t \rightarrow \infty} U({q};t)$) for several values of the bias voltage in the NDR region are depicted in Fig. 1 (b).
 
The I-V characteristics are presented in Fig. 2 (a). NDR behavior and hysteresis in the I-V curve are observed between 0.272V and 0.322V.
Moreover, two
plateau-like structures appear between 0.282V and 0.298V and
between 0.300V and 0.318V in the case of increasing bias. The time evolution of current in the NDR region for several values of the bias voltage is presented in Fig. 2 (b). 
It is seen that although the currents exhibit transient oscillation just after the bias
voltage is changed, in all cases considered here, other than that of 0.290V, this oscillatory
behavior decays after several picoseconds. 
A similar I-V curve was obtained by Jensen and Buot\cite{BuotPRL1991}. However, while they observed only one plateau region, our results indicate the existence of two plateau regions. In addition, while the oscillation they found does not decay, the oscillation found here does besides the cases in the upper plateau. 
Although multiple plateau structures have been observed experimentally,\cite{Asada98,Slight08,Meissner12} this is the first time that they have been reported in a theoretical study.


\begin{figure}[t]
 \includegraphics[width=7.5cm]{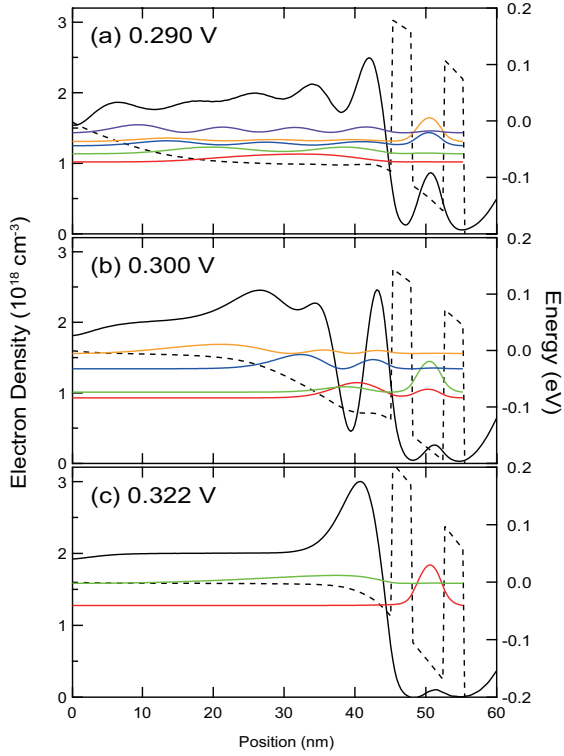}
\caption{\label{fig3:g}Time averaged electron density (black solid curve) and effective potential (black dashed curve) for (a) 0.290V, (b) 0.300V and (c)0.322V. The red, green, blue, orange, and purple curves represent the eigenfunctions in order of increasing eigenenergy calculated using the averaged effective potential without the heat bath. 
}
\end{figure}

In Fig. 1(b), we find a basin-like structure on the emitter side of each potential in the NDR region, while no such structure exists in the cases of 0.270V and 0.322V, which correspond to the upper and lower boundaries of the NDR regions, as shown in Fig. 2(a).  These emitter basins behave as a second quantum well and play an important role in the appearance of hysteresis, current oscillation and the plateau-like structure of the I-V curve\cite{FerryPRB1989, BuotJAP2000}. The basin becomes very broad in the case of 0.290V, at which value non-decaying current oscillation appears. 

To investigate the current behavior in the NDR region, we plotted the electron densities and the averaged effective potentials for several eigenstates.  These plots appear in Fig.3: (a) the oscillating 
upper plateau case (0.290V), (b) a non-oscillating lower plateau case (0.300V), and (c) the minimum case (0.322V). There, for the purpose of the graph, we employ the time-averaged potential. It should be noted, however, that in the simulations, the effective potential varies in time. In the upper plateau case, the 3rd (blue) and 4th (orange) eigenstates are the tunneling states, whereas in the lower plateau case the 1st (red) and 2nd (green) eigenstates are the tunneling states. 
Although here we did not observe a steady oscillation, the current oscillation observed in the previous studies on the single plateau case corresponds to the lower plateau case, since it arises from the ground tunneling state in the basin potential.\cite{Cui}

The frequency distribution of the current is displayed in the inset of Fig. 1(b). The lower peak there corresponds to transitions between the 5th (purple) and 4th (orange) or the 3rd (blue) and 2nd (green) eigenstates, while the higher peak can be attributed mainly to transitions between the 3rd (blue) and 1st (red) eigenstates. These results indicate that the current oscillation arises from variation of the electron distribution in the emitter basin.  The upper plateau oscillation is unique in that it arises from the excited tunneling states rather than the ground tunneling state. 
Experimentally observed multiple plateau structures\cite{Asada98, Slight08,Meissner12} may also be explained in terms of this origin.
 

\begin{figure}[t]
 \includegraphics[width=7.5cm]{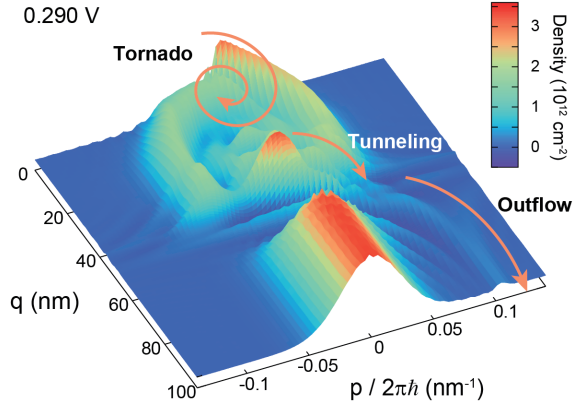}
\caption{\label{fig4:g}A snapshot (t=41.4 ps) of the Wigner distribution function for a bias voltage of 0.290V in the case of increasing bias. Current flows into the system from the emitter side of the boundary ($q$=0nm). Then, a part of the current is scattered by the emitter side of the barrier. The scattered electron flows in a tornado-like manner to central peak in the emitter basin due to dissipation. The shaking motion of the effective potential periodically accelerates the component at the central peak to the tunneling state, and the current thus exhibits steady oscillation.
}
\end{figure}

We display a snapshot of the Wigner distribution for the case of 0.290V in Fig. 4. The tornado-like distribution centered at $q$=30nm, appears in the emitter basin range between $q$=10-45nm and $p$=-0.05$-$0.05nm$^{-1}.$
This tornado-like wavepacket rotates clockwise, and whenever its tail reaches the maximum value of the momentum, near $p=0.1$nm$^{-1}$, current flows to the collector side as a result of tunneling. This phenomenon is due to the shaking motion of the effective potential, which arises from competition between the motion of the electrons and the field that they induce. 

The difference between oscillating and non-oscillating cases in the NDR region can be explained on the basis of Fig. 3. In the cases considered in Figs. 3(b) and 3(c), the tunneling states are in the bottom of the basin potential. The profiles of the electron distributions in these cases indicate that none of the populated states are the ground state, but states close to the emitter voltage. Current oscillation does not occur in these cases because the tunneling states are largely unpopulated. If the dissipation were larger, the lower tunneling state may be populated. However, due to damping, the current oscillation would be suppressed. The lower plateau feature can be explained by considering the two adjacent states at the bottom of the basin potential illustrated in Fig. 3(b). Because the other states may also result in tunneling, the output voltage does not change significantly when the emitter voltage is increased slightly. In the case of Fig. 3(c), the basin becomes so narrow that the resonant tunneling state is almost completely unpopulated. Once the bias voltage exceeds 0.322V, the basin disappears, and the I-V characteristics become normal. 

In summary, we investigated the I-V characteristics and dynamical features of the current in the NDR region on the basis of the HEOM approach, employing the Caldeira-Leggett Hamiltonian. We find that while most of the current oscillations decay in time in the NDR region, there is a steady oscillation characterized by a tornado-like rotation in the Wigner space in the upper plateau of the NDR region.
 To explore the condition of current oscillations in the NDR region, calculations for different physical conditions are necessary. We will discuss the details of these calculations in a forthcoming paper.



\clearpage
\begin{thebibliography}{10}
\bibitem{CLAnnlPhys1983} A.~O.~Caldeira and A.~J.~Leggett, Annl. Phys, {\bf 149} (1983) 374.
\bibitem{CLPhysica1983} A.~O.~Caldeira and A.~J.~Leggett, Physica, A{\bf 121} (1983) 587.
\bibitem{WaxmanLeggett1985}D. Waxman and A. J. Leggett, Phys. Rev B{\bf 32}(1985) 4450.
\bibitem{Miller1989}G. A. Voth, D. Chandler, W. H. Miller, J. Chem. Phys. {\bf 91} (1989) 7749. 
\bibitem{Chen1986}Y.-C. Chen, J. Low Temp. Phys. {\bf 65} (1986) 133.
\bibitem{MUKAMEL95}S. Mukamel,\textit{Principles of Nonlinear Optical Spectroscopy} (Oxford University Press, New York, (1995).
\bibitem{Hanggi09} P. H\"{a}nggi and F. Marchesoni, Rev. Mod. Phys.{\bf 81}(2009) 387. 
\bibitem{Waxman85}L-D. Chang and D. Waxman, J Phys C {\bf 18} (1985) 5873.
\bibitem{Coffey07}W. T. Coffey
et al, Phys. Rev. E  {\bf 75} (2007) 041117.
\bibitem{Jyoti2011}A. Shit, S. Chattopadhyay, and J. R. Chaudhuri, Chem. Phys. {\bf 386} (2011) 56.
\bibitem{TKJSPS1989} Y.~Tanimura and R.~Kubo, J. Phys. Soc. Jpn. {\bf 58} (1989) 101.
\bibitem{Tanimura91}Y. Tanimura and P. G. Wolynes, Phys. Rev. A{\bf 43} (1991) 4131;  J. Chem. Phys. {\bf 96} (1992) 8485 .
\bibitem{TJPSJ2006} Y.~Tanimura, J. Phys. Soc. Jpn. {\bf 75} (2006) 082001.
\bibitem{Datta} S. Datta, \textit{Electronic Transport in Mesoscopic Systems}, Cambridge University Press, 1995
\bibitem{FerryGoodnickBird} D. K. Ferry, S. M. Goodnick, and J. Bird, \textit{Transport in Nanostructures}, Cambridge University Press (2nd edition), 2009
\bibitem{CEsakiTsuAPL1974} L. L. Chang, L. Esaki, and R. Tsu, Appl. Phys. Lett. { \bf 24}(1974) 593 .
\bibitem{SollenerAPL1984} T. C. L. G. Sollner
et al, Appl. Phys. Lett. {\bf 45} (1984) 1319.
\bibitem{GoldmanTsuiPRL1987} V. J. Goldman, D. C. Tsui, and J. E. Cunningham, Phys. Rev. Lett. { \bf 58} (1987) 1256.
\bibitem{FrensleyPRB1987} W. R. Frensley, Phys. Rev. B {\bf 36} (1987) 1570.
\bibitem{FrensleyRevModPhys1990} W. R. Frensley, Rev. Mod. Phys. {\bf 62}(1990) 745.
\bibitem{FerryPRB1989} N. C. Kluksdahl
et al, Phys. Rev. B {\bf 39} (1989) 7720.
\bibitem{BuotPRL1991} K. L. Jensen and F. A. Buot, Phys. Rev. Lett. {\bf 66}(1991) 1078.
\bibitem{BuotJAP2000} P. Zhao
et al, J. Appl. Phys. {\bf 87} (2000) 1337.
\bibitem{ITJSPS2005} A.~Ishizaki and Y.~Tanimura, J. Phys. Soc. Jpn. { \bf 74} (2005) 3131.
\bibitem{shi09} Q. Shi
et al, J. Chem. Phys. {\bf 130} (2009) 134505.
\bibitem{STJPCA2011}A. Sakurai and Y. Tanimura, J. Phys. Chem. A {\bf 115} (2011) 4009.
\bibitem{Yan12} F. Jiang
et al, Phys. Rev. B {\bf 85} (2012) 245427.
\bibitem{IshiFlem09}A. Ishizaki and G. R. Fleming, Proc. Natl. Acad. Sci. U.S.A. {\bf 106} (2009) 17255.
\bibitem{Kramer12}C. Kreisbeck and T. Kramer, J. Phys. Chem. Lett., {\bf 3} (2012) 2828.
\bibitem{Schulten12}J. Str\"{u}mpfer and K. Schulten, J. Chem. Phys. {\bf 137} (2012) 065101.
\bibitem{TTJPSJ08}M. Tanaka and Y.~Tanimura, J. Phys. Soc. Jpn. {\bf 78} (2009) 073802;  J. Chem. Phys. {\bf  132} (2010) 214502.
\bibitem{TJCP2012}Y.~Tanimura, J. Chem. Phys. {\bf 137} (2012) 22A550.
\bibitem{Dijkstra10}A. G. Dijkstra and Y. Tanimura, Phys. Rev. Lett. {\bf 104} (2010) 250401; J. Phys. Soc. Jpn. {\bf 81}(2012) 063301.
\bibitem{Nori12} X. Yin
et al, Phys. Rev. A {\bf 86} (2012) 012308.
\bibitem{Asada98}M. Asada, S. Suzuki, and N. Kishimoto, Jpn. J. Appl. Phys. {\bf47}  (2008) 4375.
\bibitem{Slight08}T. J. Slight et al, IEEE J. Quantum Electron. {\bf44} (2008) 1158.
\bibitem{Meissner12}M. Feiginov et al, EPL {\bf 97} (2012) 58006.
\bibitem{Cui}P. Zhao, D. L. Woolard, H. L. Cui, Phys. Rev. B {\bf 67} (2003) 085312.
\end{thebibliography}
\end{document}